\documentclass{moriond}

\def\be{\begin{equation}}
\def\ee{\end{equation}}
\def\bea{\begin{eqnarray}}
\def\eea{\end{eqnarray}}

\def\SD{StarDICE}
\def\lemaitre{\textsc{Lemaître}}

\newcommand\footurl[1]{\footnote{\url{#1}}}
\def\sn{supernov\ae}
\usepackage{siunitx}

\newcommand{\Photo}{\includegraphics[height=35mm]{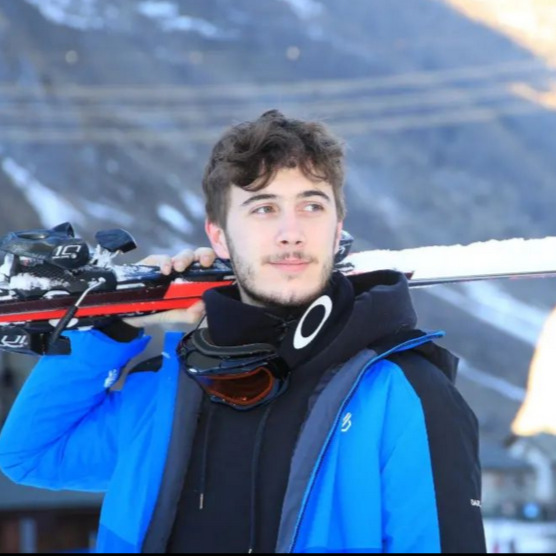}}

\begin{document}

\vspace*{4cm}
\title{HOW THE \SD{} PHOTOMETRIC CALIBRATION OF STANDARD STARS CAN IMPROVE COSMOLOGICAL CONSTRAINTS?}

\author{T.~SOUVERIN$^1$, J.~NEVEU$^{1, 3}$, M.~BETOULE$^1$, S.~BONGARD$^1$, P.~E.~BLANC$^5$, J.~COHEN~TANUGI$^{2, 6}$, S.~DAGORET-CAMPAGNE$^3$, F.~FEINSTEIN$^4$, M.~FERRARI$^5$, F.~HAZENBERG$^1$, C.~JURAMY$^1$, L.~LE~GUILLOU$^1$, A.~LE~VAN~SUU$^5$, M.~MONIEZ$^3$, E.~NUSS$^2$, B.~PLEZ$^2$, N.~REGNAULT$^1$, E.~SEPULVEDA$^1$, K.~SOMMER$^2$}

\address{$^{1}$Sorbonne Universit\'e, CNRS, Universit\'e de Paris, LPNHE, 75252 Paris Cedex 05, France; 
$^2$LUPM, Université Montpellier \& CNRS, F-34095 Montpellier, France; 
$^3$Universit\'e Paris-Saclay, CNRS, IJCLab, 91405, Orsay, France; 
$^4$CPPM, Université d'Aix-Marseille \& CNRS, 163 av. de Luminy 13288 Marseille Cedex 09, France; 
$^5$Observatoire de Haute-Provence, Université d'Aix-Marseille \& CNRS, 04870 Saint Michel L'Observatoire, France;
$^6$LPC, IN2P3/CNRS, Université Clermont Auvergne, F-63000 Clermont-Ferrand, France}

\maketitle\abstracts{
The number of type Ia supernova (SNe Ia) observations will grow significantly within the next decade, mainly thanks to the Legacy Survey of Space and Time (LSST) undertaken by the Vera Rubin Observatory in Chile. With this improvement, statistical uncertainties will decrease, and flux calibration will become the main uncertainty for the characterization of dark energy. Currently, the astronomical flux scale is anchored on the numerical models of white dwarf atmospheres from the CALSPEC catalog, and every error on the model can induce a bias over cosmological parameters inference. The \SD{} experiment proposes a new calibration reference that only relies on observations from the optical watt defined by the NIST towards the magnitude of standard stars. It is currently operating at l’Observatoire de Haute-Provence and has been collecting data since the beginning of 2023. To overcome the photometric calibration uncertainty and reach a sub-percent precision, the instrument throughput has been calibrated with a Collimated Beam Projector. It will be monitored on-site with a LED-based artificial star source calibrated with NIST photodiodes. In this proceeding, we will first illustrate how an error in the photometric calibration can impact the SNe Ia distance moduli and thus bias the measurement of cosmological parameters. Then we will present the \SD{} experiment and how we can recalibrate the CALSPEC catalog at the millimagnitude level on the NIST scale with photometric analysis.
}

\section{CALSPEC calibration}

\begin{figure}
    \centering
    \includegraphics[width=.75\textwidth]{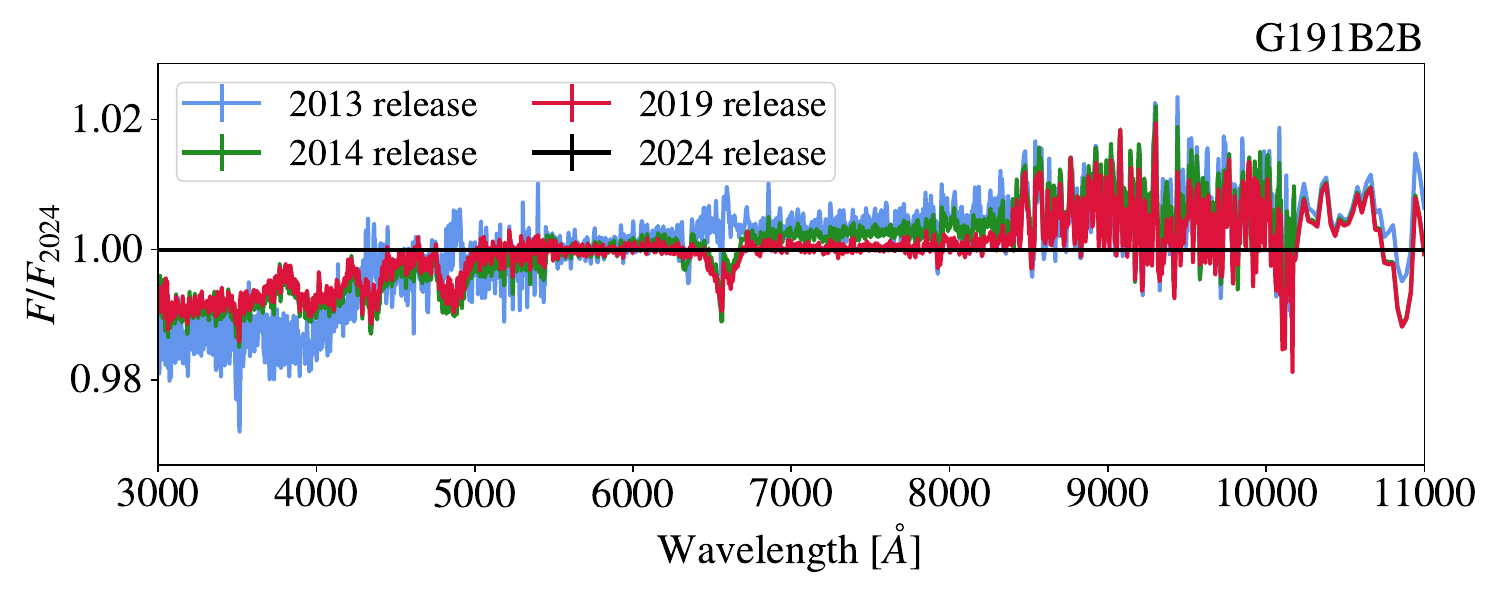}
    \caption{Relative flux of the G191B2B standard between CALSPEC releases normalized by the flux from the 2024 release and at $F(\lambda\simeq\SI{5556}{\AA})$.}
    \label{fig:calspec_diff}                
\end{figure}

Nowadays, the CALSPEC calibration is the state-of-the-art for the astrophysical photometric calibration. The three CALSPEC prime standards are pure hydrogen \textit{white dwarfs} (WD), named G191B2B, GD71 and GD153. The calibration of their \textit{spectral density energy} (SED) is anchored on numerical model of their atmosphere. In few words, the numerical radiative transfer model of their atmosphere is fitted on the hydrogen absorption lines obtained with\ high-resolution spectroscopy from the \textit{Space Telescope Imaging Spectrograph} (STIS) of the \textit{Hubble Space Telescope} (HST) (see \textbf{Bohlin et al., 2014}\cite{2014PASP..126..711B} for the detailed method). Over the past decade, the CALSPEC calibration has undergone multiple releases, coming from the evolution of the model used to simulate the WD atmosphere. Hence, the calibrated spectra of the three primary standards vary with the CALSPEC release used. Figure~\ref{fig:calspec_diff} illustrates the relative flux in the G191B2B SED, revealing a chromatic variation of up to $\sim 2\%$ between the different CALSPEC releases.

\section{Simulation of SNe Ia observation and calibration}

We aim to track variations in the calibration and their impact on dark energy parameter constraints using simulations from the \lemaitre{} project presented by \textbf{Lacroix, this proceeding}. We simulate SNe Ia light curves with the Mocksurvey\footnote{\url{https://gitlab.in2p3.fr/lemaitre/mocksurvey}}  package based on a $w$CDM model with $w=-1$, following the redshift distribution of the three surveys. The number of SNe Ia simulated are 1841 for ZTF; 886 for SNLS; and 659 for HST, reaching quantities equivalent to the \lemaitre{} combination. The simulated bandpasses are generated with the Bandpasses\footnote{\url{https://gitlab.in2p3.fr/lemaitre/bandpasses}} package. 



To calibrate an instrument, one effective method is to observe an on-sky reference like a CALSPEC standard star, which is assumed to be accurate. By comparing the expected integrated flux of the spectrum over the filter bandpass with the measured flux, we can determine an offset calibration term for each filter, known as \textit{zero points}. They serve as reference values, anchoring our measurements to a photometric system reference, such as the CALSPEC calibration.


To investigate the influence of different CALSPEC releases on the filter transmissions of our surveys, we simulate the observation of the 2024 CALSPEC release of G191B2B by integrating the \lemaitre{} survey filter transmissions. This simulation is repeated for the 2013 release, and we compute the flux difference for each filter of each survey using the two CALSPEC releases. The differences quantify the calibration bias from using one release over the other. To examine the impact of this calibration bias on the fit of cosmological parameters, we add the flux bias to the corresponding zero point value for each filter, producing a biased dataset of SNe Ia simulations. This keeps the observed flux constant while altering only the zero points used to fit the cosmological parameters.

We then generate a Hubble diagram by fitting the distance moduli $\mu$ of each SN with SNCosmo\footurl{https://github.com/nregnault/sncosmo}. Figure~\ref{fig:calspec_diff} shows the residual plot to the simulated cosmology, revealing significant scatter and an important bias for the SNLS survey. This analysis underlines that a miscalibration of approximately \SI{10}{\milli mag} on $\mu$ can lead to a bias exceeding \SI{3}{\%} in the constraint of $w$. Consequently, achieving a \SI{1}{\milli mag} accuracy in the measurement of CALSPEC standards is crucial to reach $w$ accuracy below \SI{3}{\%}, aligning with the precision goals of modern cosmology.

\begin{figure}
    \centering
    \includegraphics[width=1\textwidth]{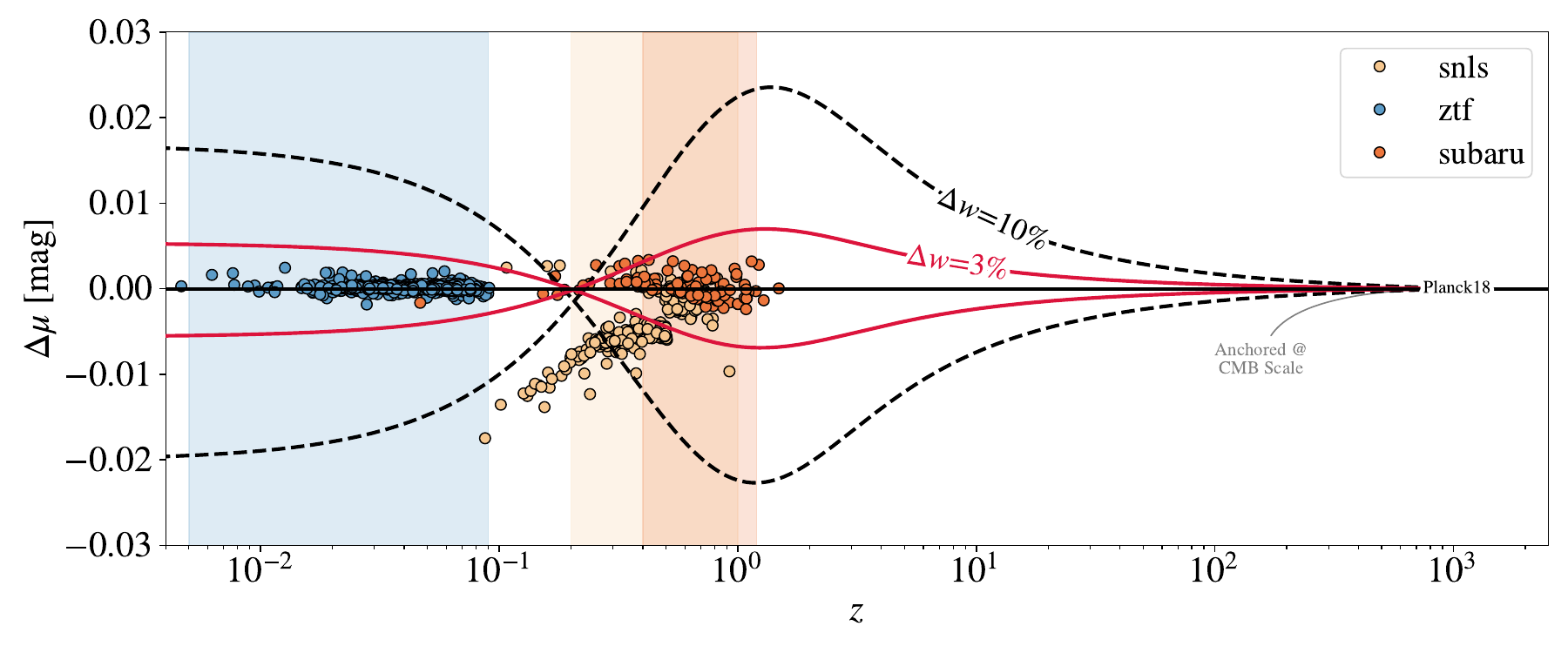}
    \caption[Hubble diagram residuals for \lemaitre{}.]{Residuals of a Hubble diagram modified by introducing a flux calibration bias in the simulation. The red and black dashed lines represent how a deviation of $w$ of respectively \SI{3}{\%} and \SI{10}{\%} impacts the distance moduli $\mu$. The colored surfaces represent an approximate visualization of the redshifts ranges of each survey.}
    \label{fig:w0_sens_calspec}
\end{figure}


\section{\SD{} experiment}

An approach to meet this requirement involves directly measuring the absolute flux of CALSPEC spectra without relying on prior numerical models. This is achieved by transferring the calibration from laboratory standards to observations of CALSPEC standards, as discussed in Section 6 of \textbf{Bohlin et al. 2014}\cite{2014PASP..126..711B}. 

\begin{figure}
    \centering
    \includegraphics[width=1\textwidth]{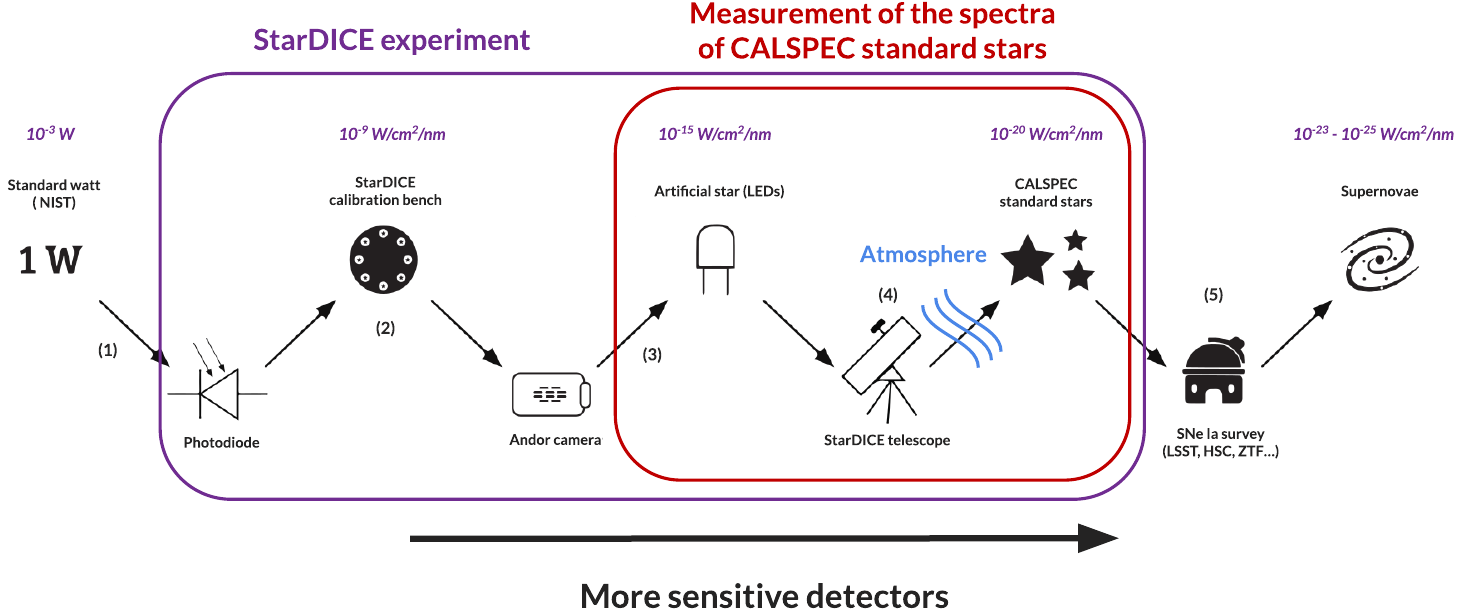}
    \caption[\SD{} calibration transfer.]{Calibration transfer of the \SD{} experiment, from the standard watt defined by the NIST to the measurement of a type Ia \sn{}. The top row corresponds to the sources, while the bottom shows the detectors. The purple box represents every step undertaken by the \SD{} experiment, while the red box shows the calibration of the standard stars. Every step of this transfer aims to calibrate more sensitive detectors and goes through approximately 20 orders of magnitude of luminosity.}
    \label{fig:sd_transfer}
\end{figure}

The \SD{} experiment proposes a metrology chain from laboratory flux references toward the measurement of standard star spectra. This calibration transfer is schematized in Figure~\ref{fig:sd_transfer}, and goes as follows: (1) is operated
by the NIST (\textbf{Houston et al. 2008}\cite{1197111}) and results in a silicon photodiode calibrated against the POWR\footnote{\url{https://www.nist.gov/laboratories/tools-instruments/primary-optical-watt-radiometer-powr-facility}} facility; (2) is achieved with the \SD{} sensor calibration bench (\textbf{Betoule et al. 2023}\cite{2023A&A...670A.119B}); (3) consists of the calibration of a \SD{} artificial star with the previously calibrated \SD{} camera; (4) is the recalibration of the CALSPEC standard stars with the calibrated \SD{} telescope (\textbf{Souverin et al. 2024 in prep.}); (5) is performed by the survey itself, observing standard stars to calibrate its instrument.


The \SD{} photometric instrument is a Newton telescope hosted l'Observatoire de Haute-Provence, observing in the \textit{ugrizy} photometric system. It has been operational since early 2023, currently undergoing a pre-survey phase focused on testing observing strategies. Since September 2023, it has cumulated around 30 nights of observations. During this phase, the priority has been given to observing CALSPEC primary standards when they are visible, and secondary standards otherwise, with several targets observed each night. A first photometric analysis have been pursued to evaluate the \SD{} performance to refine the CALSPEC calibration, for every $ugrizy$ filter. We have analyzed a total of 9 observation nights of G191B2B, and studied the accuracy at which \SD{} can measure the zero point of each filter. 

\begin{figure}
    \centering
    \includegraphics[width=1\textwidth]{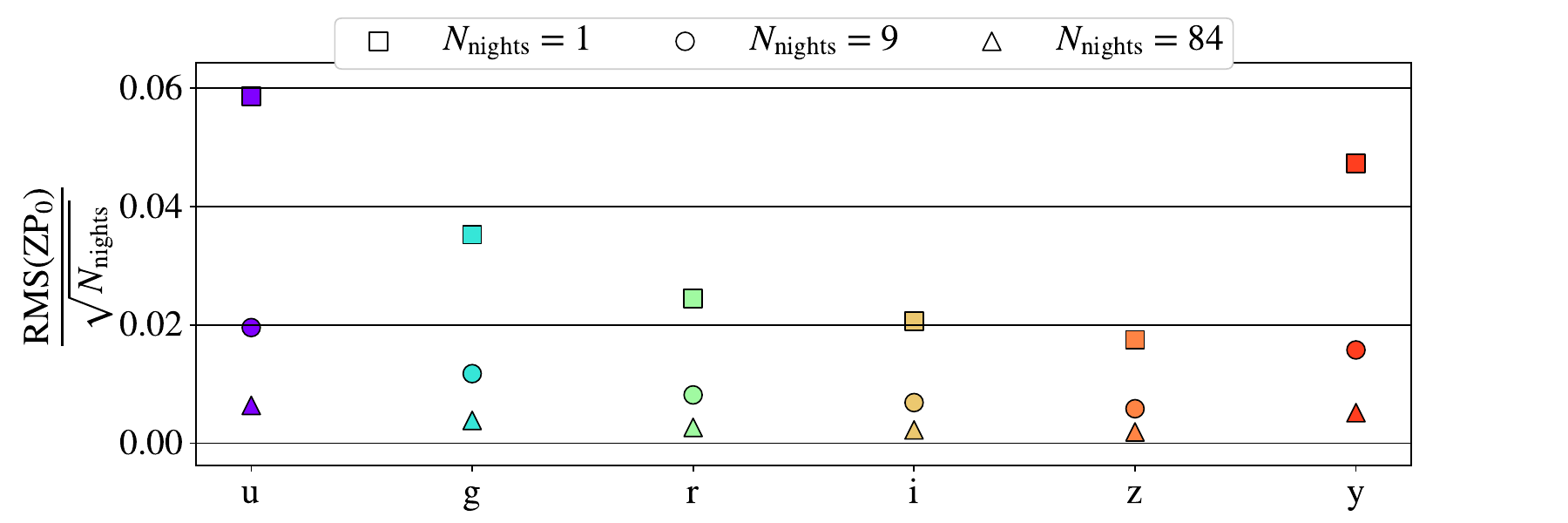}
    \caption{RMS of $\mathrm{ZP}_0$ fitted for the 9 photometric nights of observations and for each \SD{} band, normalized by the square root of the number of nights. The square represents the RMS for one night of observations; the circles represent the RMS for the 9 photometric nights used in this analysis; the triangles represent the predicted RMS for 84 observation nights, corresponding to approximately 2 years of the \SD{} survey. }
    \label{fig:sd_results}
\end{figure}

The results are depicted in Figure~\ref{fig:sd_results}, showing the \textit{Root Mean Square} of the zero point measurement for each filter, normalized by the square root of the observation nights $N_\mathrm{nights}$. With $N_\mathrm{nights} = 9$, the \SD{} experiment demonstrates the ability to refine the CALSPEC calibration to a dispersion of about \SI{1}{\%} for the $griz$ bands. Scaling this performance to a  2-year survey, equivalent to $\sim$ 80 observation nights, we estimate achieving a dispersion of \SIrange{0.2}{0.4}{\%} for the $griz$ bands. This precision is roughly two to four times higher than the target needed to reduce the systematic bias on the $w$ determination to less than \SI{3}{\%}. 

Significant improvements are expected for the \SD{} experiment in the coming years, which will hopefully reduce the zero point dispersion. These enhancements include the installation of an artificial star to continuously monitor telescope calibration, an infrared camera to track cloud-induced extinction, and the implementation of spectrophotometric analysis for more accurate atmospheric contribution estimations.

\end{document}